\pdfoutput=1
\documentclass[fleqn,usenatbib]{mnras}

\usepackage{newtxtext,newtxmath}

\usepackage[T1]{fontenc}
\usepackage{ae,aecompl}
\usepackage{soul}
\usepackage{aas_macros}
\usepackage[caption=false]{subfig}
\usepackage{ae,aecompl}
\usepackage{hyperref}
\usepackage{graphicx}
\usepackage{epsf}
\usepackage{longtable}
\usepackage{rotating}
\usepackage{bm}
\usepackage{color}
\usepackage{amsmath}

\usepackage{booktabs}
\usepackage{braket}
\usepackage{multirow}
\setcitestyle{square}
\usepackage{xcolor}
\usepackage{etoolbox}
\makeatletter
\patchcmd\@combinedblfloats{\box\@outputbox}{\unvbox\@outputbox}{}{\errmessage{\noexpand patch failed}}
\makeatother

\title{Weak Lensing Minima and Peaks: Cosmological Constraints and the Impact of Baryons}
\author[W.R Coulton et al.]{
William R. Coulton,$^{1}$\thanks{E-mail: wcoulton@ast.cam.ac.uk}
Jia Liu,$^{2,3,4}$
Ian G. McCarthy$^{5}$
and Ken Osato$^{6}$
\\
$^{1}$Institute of Astronomy and Kavli Institute for Cosmology Cambridge, Madingley Road, Cambridge, CB3 0HA, UK\\
$^{2}$Berkeley Center for Cosmological Physics, University of California, Berkeley, CA 94720, USA\\
$^{3}$Lawrence Berkeley National Laboratory, 1 Cyclotron Road, Berkeley, CA 93720, USA \\
$^{4}$Department of Astrophysical Sciences, Princeton University, Peyton Hall, Princeton, NJ 08544, USA \\
$^{5}$Astrophysics Research Institute, Liverpool John Moores University, 146 Brownlow Hill, Liverpool, L3 5RF, UK \\
$^{6}$Institut d'Astrophysique de Paris, Sorbonne Universit\'e, CNRS, UMR 7095, 98bis boulevard Arago, F-75014 Paris, France
}
\begin{document}
\label{firstpage}
\pagerange{\pageref{firstpage}--\pageref{lastpage}}
\maketitle
\begin{abstract}
We present a novel statistic to extract cosmological information in weak lensing data: the lensing minima. We also investigate the effect of baryons on the cosmological constraints from peak and minimum counts. 
Using the \texttt{MassiveNuS} simulations, we find that lensing minima are sensitive to  non-Gaussian cosmological information and are complementary to the lensing power spectrum and peak counts. For an LSST-like survey, we obtain $95\%$ credible intervals from a combination of lensing minima and peaks that are significantly stronger than from the power spectrum alone, by $44\%$, $11\%$, and $63\%$ for the neutrino mass sum $\sum m_\nu$,  matter density $\Omega_m$, and amplitude of fluctuation $A_s$, respectively. We explore the effect of baryonic processes on lensing  minima and peaks using the hydrodynamical simulations \texttt{BAHAMAS} and \texttt{Osato15}. We find that ignoring baryonic effects would lead to strong ($\approx 4 \sigma$) biases in inferences from peak counts, but negligible ($\approx 0.5 \sigma$) for minimum counts, suggesting lensing minima are a potentially more robust tool against baryonic effects.  Finally, we demonstrate that the biases can in principle be mitigated without significantly degrading cosmological constraints when we model and marginalize the baryonic effects.   
\end{abstract}

\section{Introduction}
Weak gravitational lensing of background galaxies  probes the integrated matter fluctuations along the line-of-sight. It is sensitive to fundamental physics such as the nature of dark energy and the total mass of neutrinos~(see a recent review by~\cite{Kilbinger2015}). Recently, pioneering weak lensing surveys achieved high precision  measurements, 
leading to competitive constraints on cosmology~\cite{heymans2012,2017DES,Hildebrandt2017,Hikage2019}. 
At present, results from weak lensing measurements have been primarily driven by two-point (or second-order) measurements, such as the two-point correlation function or its Fourier transformation, the power spectrum. However, nonlinear structure growth on scales smaller than a few$\times 10 \, \mathrm{Mpc}$ moves a significant portion of cosmological information from second-order to higher-order statistics.  As the next generation weak lensing surveys are coming online in the next decade, such as the LSST\footnote{Large Synoptic Survey Telescope: \url{http://www.lsst.org}}~\cite{LSSTscienceBook}, WFIRST\footnote{Wide-Field Infrared Survey Telescope: \url{http://wfirst.gsfc.nasa.gov} }, and \textit{Euclid}\footnote{\textit{Euclid}: \url{http://sci.esa.int/euclid} }, we are probing deep into the nonlinear regime and two-point statistics will no longer be adequate. 

To retrieve the lost information, we need to utilize higher-order, non-Gaussian statistics. Naturally, one would consider the next-order term, the three-point correlation function (or its Fourier transformation, the bispectrum)~\cite{Takada2003,Vafaei2010,fu2014,Coulton2019}. However, three-point functions can be computationally expensive, due to the large number of possible triangle shapes in a typical lensing dataset, and have large, complex covariance matrices. It also has the drawback of only capturing third-order information, neglecting fourth- and higher-order information. Therefore, weak lensing summary statistics, which can be easier to measure and also contain information of all orders, have been proposed as simpler alternatives, such as the peak counts~\cite{Jain2000b,Marian2009,Maturi2010,Yang2011,Marian+2013,Liu2015,Liu2014b,Lin&Kilbinger2015a,Lin&Kilbinger2015b,Kacprzak2016,Peel2018,Shan2018,Martinet2018,Li2019}, Minkowski functionals~\cite{Kratochvil2012,Shirasakiyoshida2014,Petri2013,Petri2015,Marques2019}, and higher-order moments~\cite{Bernardeau1997,Hui1999,vanWaerbeke2001, Takada2002,Zaldarriaga2003,Kilbinger2005,Petri2015,Peel2018}.

In this paper, we propose a new probe of non-Gaussian information --- weak lensing minima. 
The simple name already describes well the nature of the statistic --- the number counts of local minima in a (typically smoothed) lensing convergence map as a function of their depth. Computationally, they are the pixels with values smaller than their surrounding pixels. Our motivation to investigate lensing minima is threefold. First, they probe the emptiest regions (voids) in our universe, and hence are complementary to the well-investigated lensing peaks which are typically associated with massive halos.  Second, the baryonic effects are expected to impact voids differently than overdense regions, 
leading us to postulate that baryon physics, one of the most worrisome lensing systematics, would impact lensing minima differently than other non-Gaussian statistics and may in turn help constrain baryon physics. Finally, lensing minima are a particularly simple non-Gaussian statistic, as they are easily computed from observational data and can be modeled using existing weak lensing simulations built for other lensing statistics. 

Lensing signals around underdense regions in the universe have been previously studied both theoretically and with observational data. For example, \cite{Sanchez2017} used DES redMaGiC galaxies to identify voids and found the stacked lensing signal around them to be negative at a $4\sigma$ level (also see~\cite{Melchior2014,Clampitt2015,Gruen2016}). The lensing profile around underdense regions has also been a topic of interest for modified gravity, as dark energy is more prominent in void regions~\cite{Barreira2015,Barreira2017,Baker2018,Paillas2019}. 
Void lensing can be more complicated than cluster lensing, as voids may have irregular shapes and the lensing signal depends on the void identification scheme. For example, \cite{Davies2018}  found a higher lensing signal for voids identified using lensing peaks, compared to those found using galaxies. 

Our method has the advantage of simplicity --- lensing minima do not require any void tracer, such as halos or lensing peaks as needed in previous works. While we expect the lensing minima to be mostly associated with void regions (i.e. with negative lensing signal), we find that a small number of them have positive values and hence also have imprints from slightly overdense regions. In this paper, we  show that lensing minima are sensitive to cosmology and forecast the cosmological constraints for an LSST-like survey. We compare our results to the power spectrum, a traditional lensing measurement, and the peak counts, a more well-studied lensing non-Gaussian statistic.

Measurements of peak and minimum counts can be sensitive to very small scales. As such, they could be impacted by the effects of baryons. Baryonic processes result in a suppression of the matter power spectrum at scales $k\approx 1 \, h \, \mathrm{Mpc}^{-1}$, mainly due to 
feedback processes of active galactic nuclei (AGN) and supernova explosions, and an enhancement of the very small scale power spectrum $k > 10 \, h \, \mathrm{Mpc}^{-1}$ due to baryonic cooling~\cite{vandaalen2011,Schaye2015,McCarthy2017,Nelson2019}. These processes may impact peak and minimum counts by altering the depth of the potential wells. 
The impact of baryonic processes on peak counts has been studied in previous works~\citep[e.g.][]{yang2013,Osato2015,Fong2019,Weiss2019}. Here, we extend the study to cosmological parameter constraints, using hydrodynamical simulations. We will also present the first study of the impact of baryons on lensing minima.

The paper is organized as follows. In Section~\ref{sec:simulations}, we provide an overview of the simulations used in this work. In Section~\ref{sec:minimaCounts}, we explore the properties of lensing minima measured in our simulated lensing convergence maps and compare them with power spectrum and peak counts. We then explore the cosmological constraints from lensing minima in Section~\ref{sec:constraints} and compare  to that from the power spectrum and peak counts for an LSST-like survey. We also show that baryon physics have different impact on minima compared to the other two statistics in Section~\ref{sec:baryon}.
Finally, we conclude in Section~\ref{sec:conclusions}.

\section{Simulations}\label{sec:simulations}

To understand the cosmological power of lensing minima and their sensitivity to baryon physics, we use two types of simulations --- a set of dark-matter only simulations that model a grid of cosmological parameters, the Cosmological Massive Neutrino Simulations (\texttt{MassiveNuS}~\cite{Liu2018MassiveNuS:Simulations}) and two sets of hydrodynamical simulations --- the BAryons and HAloes of MAssive Systems (\texttt{BAHAMAS}~\cite{McCarthy2017,McCarthy2018}) and the set used in ~\cite{Osato2015} (\texttt{Osato15}). Here we briefly describe the key aspects of these simulations relevant to this work, and refer the readers to the code papers for more details.

\texttt{MassiveNuS} consists of 101 flat-$\Lambda$-cold dark matter (CDM) $N$-body simulations, with three varied parameters: the neutrino mass sum $\sum m_\nu$, the total matter density $\Omega_m$, and the  amplitude of primordial fluctuation $A_s$, covering the parameter ranges $\sum m_\nu$=[0, 0.62]~eV, $\Omega_m$= [0.18, 0.42], and $10^9 A_s$= [1.29, 2.91]. The simulations have a box size of $512 \, h^{-1} \, \mathrm{Mpc}$ and $1024^3$ CDM particles, accurately capturing structure growth at $k<10 \, h \, \mathrm{Mpc}^{-1}$. Massive neutrinos are treated using linear perturbation theory and their clustering is sourced by the full nonlinear matter density~\citep{AB2013,Bird2018}, and the resulting accuracy of the total matter power spectrum is tested to agree with particle neutrino simulations to within $0.2\%$ for $\sum m_\nu < 0.6 \, \mathrm{eV}$.  In this work we choose one simulation with cosmological parameters $\sum m_\nu$=0.1~eV, $\Omega_m$=0.3, and $A_s$= $2.1 \times 10^{-9}$ as our fiducial cosmology. 
Weak lensing convergence maps are available for five delta-function source redshifts $z_s = 0.5, 1.0, 1.5, 2.0, 2.5$ with 10,000 realizations generated per model per source redshift using the ray-tracing code \texttt{LensTools}~\cite{Petri2016Lenstools}\footnote{\url{https://pypi.python.org/pypi/lenstools}}. Maps have $512^2$~pixels and are $3.5^2 = 12.25 \, \mathrm{deg}^2$ in size. The maps at different source redshifts are ray-traced through the same large-scale structure, and hence are properly correlated. 

To study the effect of baryons, we use two sets of hydrodynamical simulations. The \texttt{BAHAMAS} simulations have a box size of $400 \,  h^{-1} \, \mathrm{Mpc}$ and $2 \times 1024^3$ particles.  The simulations were run with the Gadget-3 TreePM SPH code, which was modified to include subgrid prescriptions for metal-dependent radiative cooling, star formation, stellar and chemical evolution, black hole formation and merging, and stellar and AGN feedback, originally developed for the OWLS project \citep{Schaye2010}.  For the fiducial \texttt{BAHAMAS} model, the parameters characterising the efficiencies of stellar and AGN feedback were adjusted to approximately reproduce the observed present-day galaxy stellar mass function (above $10^{10}$ M$_\odot$) and the amplitude of the local hot gas fraction--halo mass relation, as inferred from high-resolution X-ray observations.
In addition to the fiducial \texttt{BAHAMAS} simulation with a feedback model designed to best match these observations, we also include two additional simulations where the AGN heating temperature is raised and lowered by $0.2 \, \mathrm{dex}$, hereafter ``high AGN'' and ``low AGN''.  Varying the heating temperature in this way retains a good match to the galaxy stellar mass function but changes the gas fractions of haloes (and therefore their lensing signals) so that the simulations skirt the upper and lower bounds of the observed gas fractions.  We generate 10,000 convergence maps from 25 independent light cones \cite{McCarthy2018}, using a similar technique as in the \texttt{MassiveNuS}.

The \texttt{Osato15} simulations have a box size of $240 \, h^{-1} \, \mathrm{Mpc}$ and $2 \times 512^3$ particles.
This set of simulations employs the recipe of galaxy formation physics developed in ~\cite{Okamoto2014}, where basic baryonic processes, e.g., star formation and radiative cooling, are implemented as the subgrid model. In this model, formation and evolution of black holes is not fully traced but an ad-hoc modeling is adopted to mimic the feedback effect. At each time step, the velocity dispersion within a halo is evaluated. If the velocity dispersion exceeds the threshold value, radiative cooling within the halo is manually stopped. Thus, further star formation is suppressed.
100 convergence maps are generated for both the dark matter-only simulations and the hydrodynamical runs (referred to as ``FE'' in the original paper) based on the ray-tracing technique. 

Both sets of hydrodynamical simulations have parallel dark matter-only and hydrodynamical runs with the same initial conditions, at the WMAP 9-yr cosmology with $\{ \Omega_m,  \Omega_b,  \Omega_\Lambda, \sigma_8, n_s , h \} = \{ 0.2793, 0.0463, 0.7207, 0.821, 0.972, 0.700 \}$ \cite{hinshaw2013}. 
 The main difference between these two simulation suites is the implementation of AGN feedback. For \texttt{BAHAMAS}, the modeling of black holes is based on ~\cite{Booth2009}, where AGN feedback is modelled by thermally coupling a fraction of the rest-mass energy of the accreted gas into the surrounding medium, but for \texttt{Osato2015} star formation is simply shut down for halos with large velocity dispersion. The latter model leads to weaker feedback than observations and smaller amounts of ejected gas (which has been found by ~\cite{vanDaalen2019} to be strongly correlated with the effects of baryons on the power spectrum). By including two sets of hydrodynamical simulations, we demonstrate the differences among existing feedback models and hence stress the importance of careful modeling of baryonic effects.

\begin{table}
\centering
\begin{tabular}{l|c|c|c|c|c}
\hline
$z_s$ & $0.5$ & $1.0$ & $1.5$ & $2.0$ & $2.5$ \\
\hline
$\bar{n}_{\mathrm{gal}}$  ($\mathrm{arcmin}^{-2}$) & $8.83$ & $13.25$ & $11.15$ & $7.36$ & $4.26$ \\
\hline
\end{tabular}
\caption{\label{tab:nGalPerZ} The projected source counts per $\mathrm{arcmin}^2$ used in our tomography analysis. For the single redshift $z_s = 1$ analysis, we use $\bar{n}_{\mathrm{gal}} = 44.8 \, \mathrm{arcmin}^{-2}$.}\end{table}

To create LSST-like mocks, we follow the photometric redshift distribution and galaxy number density estimated in the LSST Science Book~\cite{LSSTSci}. We use the number density for each source redshift in Table~\ref{tab:nGalPerZ}. We then add Gaussian noise to the simulated convergence maps with a variance $V_{\rm noise}$ of
\begin{equation}
V_{\rm noise}= \frac{\sigma_s^2}{n_g},
\end{equation}
where $\sigma_s = 0.3$ is the intrinsic shape noise and $n_g$ is the galaxy number density in Table~\ref{tab:nGalPerZ}. The shape noise leads to a large number of false minima. Therefore, before measuring the minima, we smooth our maps with a Gaussian filter of $\theta_G = 2 \, \mathrm{arcmin}$. We measure the counts of minima as a function of the depth, normalized by the standard deviation of the smoothed noise-only maps $\sigma_n$, where
\begin{equation}
\sigma^2_n= \frac{\sigma_s^2}{4 \log(2) \pi \theta_G^2 n_g}.
\end{equation}

We study the cosmological constraints from minima with two different redshift settings --- the tomographic setting as shown in Table~\ref{tab:nGalPerZ} and a single redshift distribution assuming all galaxies are at $z_s = 1$. Both settings have the same total number of galaxies per $\mathrm{arcmin}^2$. Our study of baryon physics is only applied to the single redshift setting.

\section{Weak Lensing Minima}\label{sec:minimaCounts}


We identify lensing minima in our simulated convergence maps as pixels with lower values than their 8 neighbours. These minima are then binned by their depth, forming the minimum counts.  The convergence maps are smoothed first, to reduce the impact of galaxy noise. The dependence of our results on smoothing scales is explored in Appendix~\ref{sec:smoothScales}.  

\subsection{Non-Gaussian information in lensing minima}

\begin{figure*}
  \centering
    \includegraphics[width=.8\textwidth]{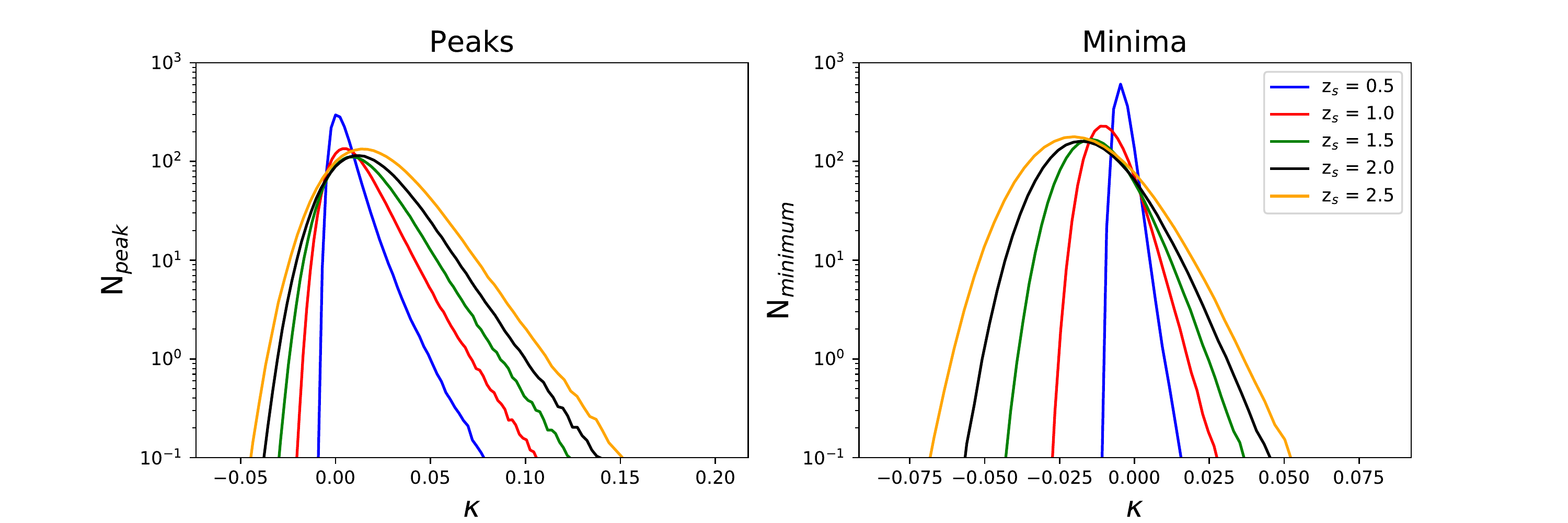}
\caption{Weak lensing peak counts (left) and minimum counts (right) as a function of the convergence $\kappa$ height/depth from the \texttt{MassiveNuS} noise-free simulations, for five source redshifts.}
\label{fig:countsNoiseFree}
\end{figure*}

In Fig.~\ref{fig:countsNoiseFree}, we show the minimum counts from the noise-free maps at the \texttt{MassiveNuS} massless fiducial cosmology, for five source redshifts. For comparison, we also show the peak counts. As expected, the minima primarily occur in underdense regions with negative $\kappa$ values. Compared to the shape of peak counts, that of the minimum counts are slightly narrower and more symmetric.  While the high $\kappa$ tails in the peak counts are formed due to the highly nonlinear regions, very long negative $\kappa$ tails in the minimum counts is absent because of a minimum possible $\kappa$ due to the density contrast limit $\delta = \rho / \bar\rho - 1 \in [-1 , +\infty)$.

\begin{figure*}
  \centering
    \includegraphics[width=.8\textwidth]{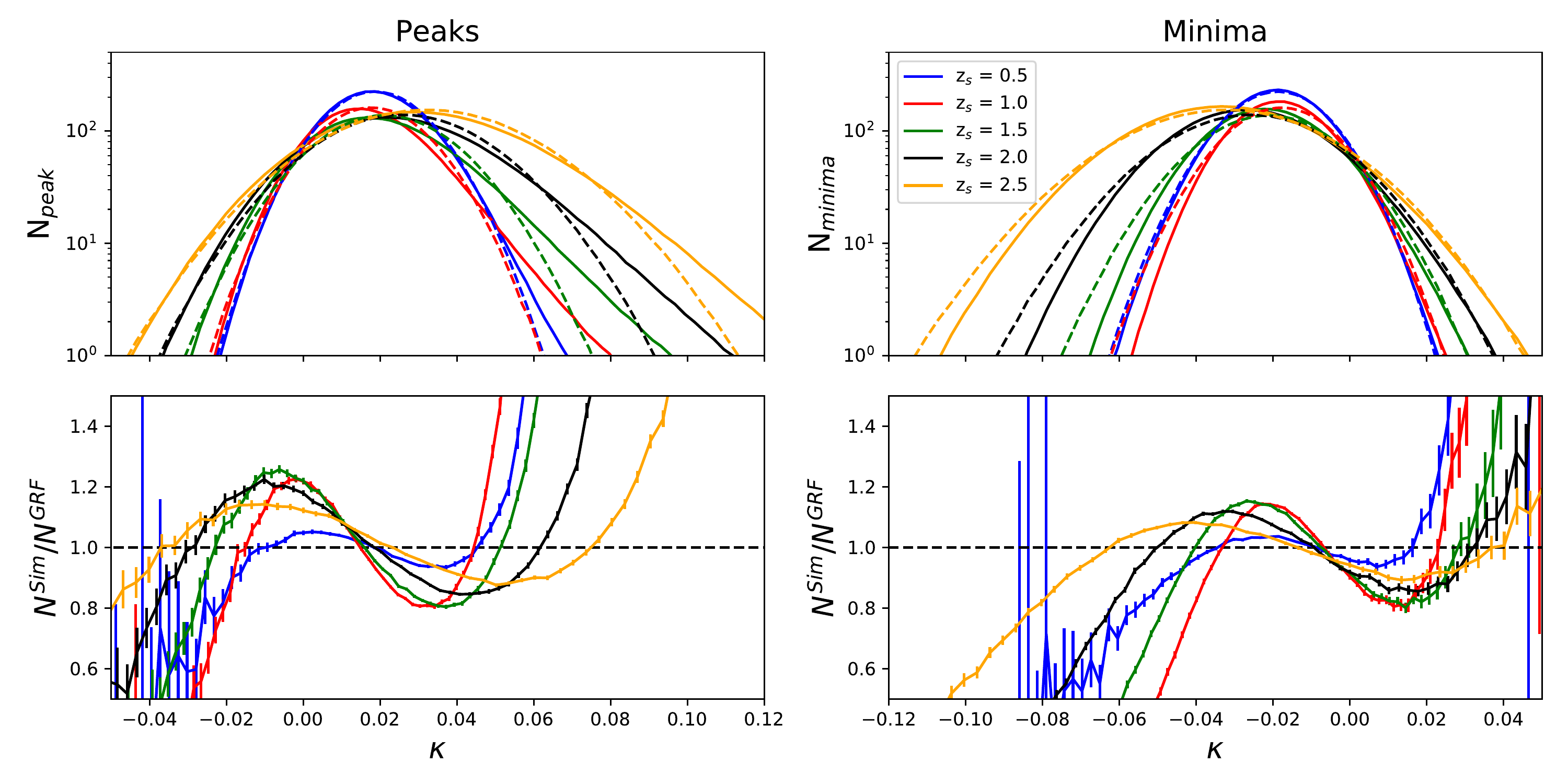}
\caption{{\bf Upper panels}: Peak counts (left) and minimum counts (right) as a function of the convergence $\kappa$ height/depth from the \texttt{MassiveNuS} simulation ({\bf solid lines}) and from Gaussian random fields (GRFs, {\bf dashed lines}). {\bf Lower panels}: ratios of the peak and minimum counts obtained from simulations to that of the GRFs. Galaxy noise is included. The error bars are for an LSST-like survey that covers 20,000 deg$^2$.}
\label{fig:countsWithNoise}
\end{figure*}

We next investigate if the non-Gaussian signals in the noise-free case remain after adding realistic galaxy noises. In Fig.~\ref{fig:countsWithNoise}, we show the number counts for maps with galaxy noises expected from LSST. In addition, we show the number counts from Gaussian random fields (GRF) that have the same power spectra as the simulated maps~\cite{Bond1987}.
For peak counts, we find more peaks at the high $\kappa$ tails in the simulations than in the GRFs, due to the nonlinear growth in the simulations beyond the second order. This is consistent with previous works on peak counts~\cite{Yang2011,Li2019}. For minimum counts, the non-Gaussian signatures are more prominent in the negative $\kappa$ tails, with less minima in the simulations than in the GRFs. 

It is clear that lensing minima contain rich non-Gaussian information that is beyond the power spectrum, even when galaxy noise is included. We next explore their sensitivity to cosmological parameters, with a focus on the sum of neutrino masses.

\subsection{Effect of neutrino mass}

\begin{figure*}
  \centering
    \includegraphics[width=.8\textwidth]{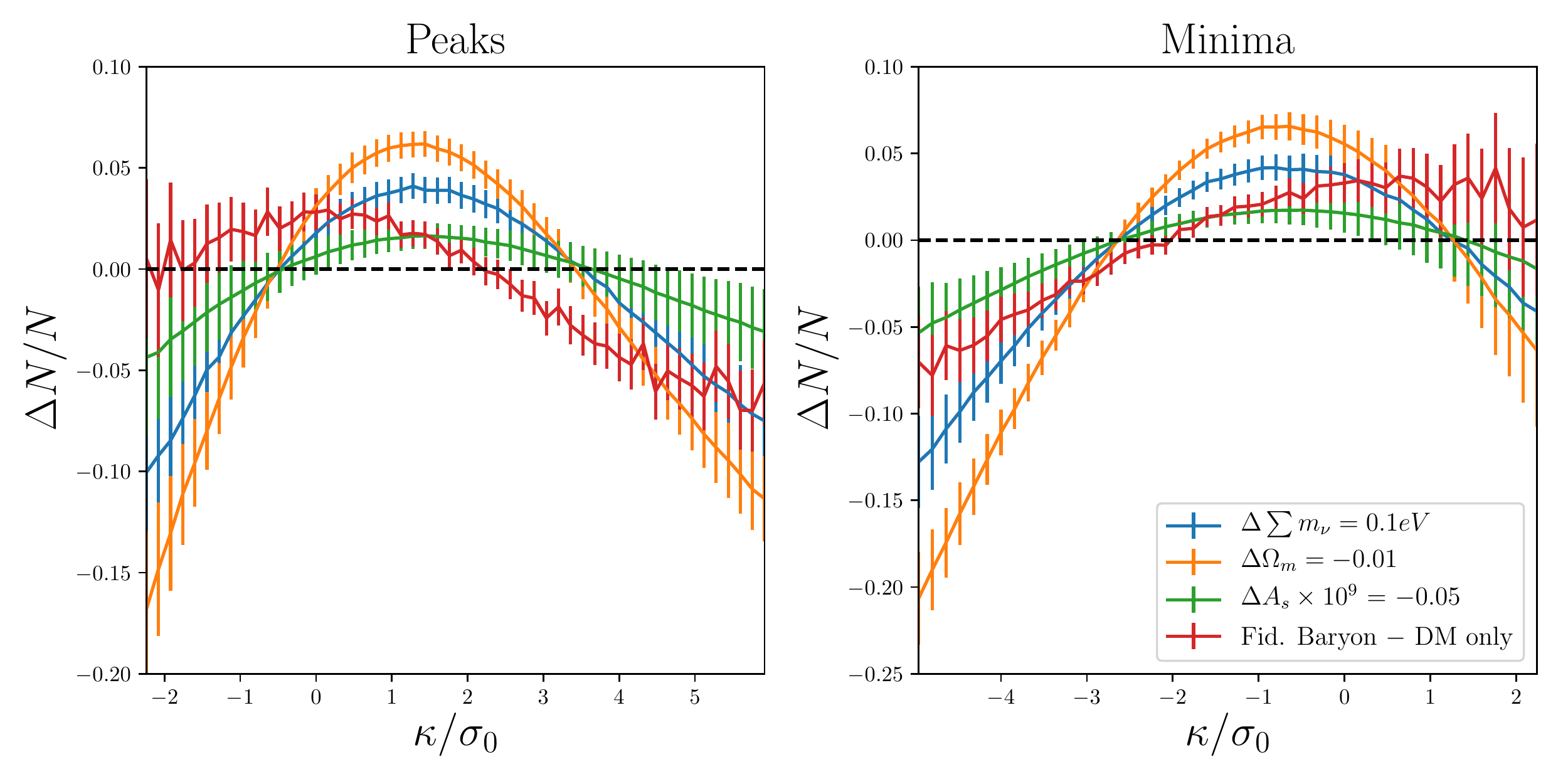}
\caption{The fractional difference in peak counts (left) and minimum counts (right) due to variation of three cosmological parameters ($\sum m_\nu$, $\Omega_m$, $A_s$) and baryons for simulations  \textbf{with LSST shape noise}, for a single source redshift $z_s=1$.  For the baryonic effects, we use the fiducial AGN feedback model and the dark matter-only model from the \texttt{BAHAMAS} simulations.} 
\label{fig:effectOfNeutrinoMassWithNoise}
\end{figure*}

Massive neutrinos suppress the growth of structures below  the free streaming scale
\begin{align}
k_F = 0.0072\left( \frac{\sum m_\nu}{0.1 \, \mathrm{eV}}\right)^\frac{1}{2}
\left(\frac{\Omega_m}{0.315}\right)^\frac{1}{2} \, h \, \mathrm{Mpc}^{-1}.
\end{align} 
We show in Fig.~\ref{fig:effectOfNeutrinoMassWithNoise} the effect of massive neutrinos, $\Omega_m$, and $A_s$ on lensing minima, by comparing a new cosmology where we vary one parameter at a time to the fiducial massive neutrino cosmology ($\sum m_\nu = 0.1 \, \mathrm{eV}$, $\Omega_m = 0.3$, $A_s \times 10^9 = 2.1$). LSST noise is included.  

When increasing the neutrino mass, we see a reduction of deep (very negative $\kappa$) minima. As neutrinos free stream from overdense regions, they ``fill in'' underdense regions, resulting in less-empty voids. Work by \cite{Massara2015,Kreisch2019} found that increasing neutrino mass reduces the number of large voids, which are necessary to create the deepest minima in lensing maps. Changes in $\Omega_m$ and $A_s$ have a similar effect, and hence can mimic the effects of neutrino mass. However, in the next section, we show that there are subtle differences in these curves that can help break the degeneracies. 

We include peak counts in Fig.~\ref{fig:effectOfNeutrinoMassWithNoise} for an easy comparison. Previously, \cite{Li2019} studied the impact of massive neutrinos on lensing peaks and found a reduction in high lensing peaks, consistent with the expectation that massive neutrinos suppress the formation of massive halos. 

We also show the effect of baryons in Fig.~\ref{fig:effectOfNeutrinoMassWithNoise}. Here we use the fiducial AGN model and dark matter-only model from the \texttt{BAHAMAS} simulations. While baryon feedback also seems to suppress deep minima, the zero-crossing points and shapes are different from those caused by cosmological parameters, for both minimum counts and peak counts. We discuss the implications in detail in Section~\ref{sec:baryon}.


\section{Cosmological Constraints}\label{sec:constraints}

To study the cosmological information in lensing minima, we use the full 101 cosmologies from the \texttt{MassiveNuS} simulation to build an emulator that models the statistics. We then use a Gaussian likelihood to estimate the credible regions for the minimum counts,  peak counts, power spectrum, and the combination of the minima and peaks. For both minimum and peak counts, we use linear signal-to-noise ratio (SNR) bins with width $\Delta \mathrm{SNR} = 0.16$ in the range $\mathrm{SNR}=[-4.16, 1.6]$ and $[-1.44, 4.48]$ for minimum counts (36 bins) and peak counts (37 bins), respectively.

\subsection{Covariance matrix}

\begin{figure*}
  \centering
    \includegraphics[width=.8\textwidth]{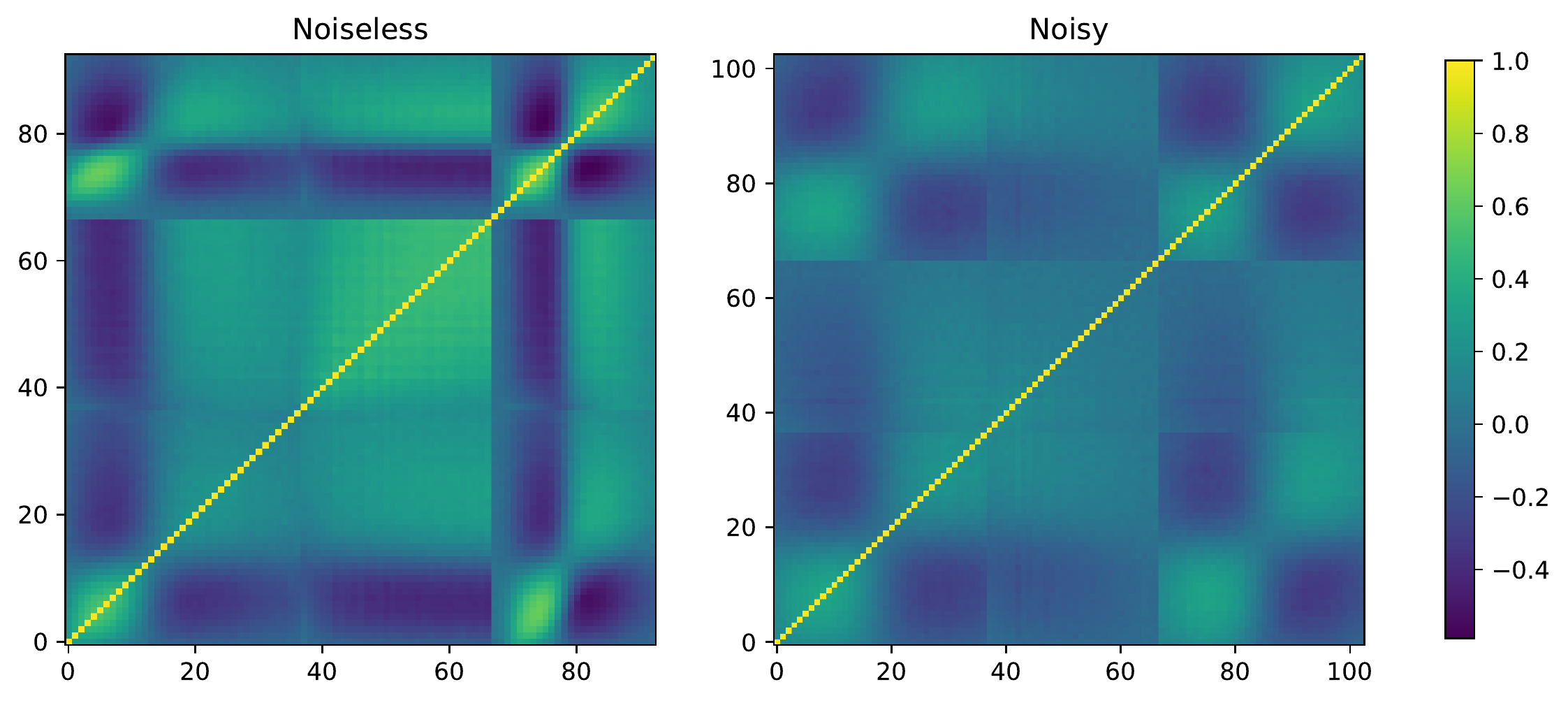}
\caption{The full covariance matrix of the peak counts (bins 1--37), power spectrum (bins 38--67), and minimum counts (left: bins 68-93; right: bins 68--103) for $z_s = 1.0$ from noiseless (left) and noisy (right) simulations. There are significantly more minima bins in the noisy case as the Gaussian noise broadens the distribution (see Figs.~\ref{fig:countsNoiseFree} and \ref{fig:countsWithNoise}).}
\label{fig:covMatThreeStats}
\end{figure*}

We explore the correlation between the minimum counts, peak counts, and power spectrum  in Fig.~\ref{fig:covMatThreeStats}, which is also the full covariance in our likelihood analysis. We use 10,000 map realizations at the fiducial cosmology to construct the covariance matrix. 
In the noise-free case, we see large off-diagonal terms for both minimum and peak counts, within their self-blocks. In addition, we also see strong correlations between the peak and minimum counts. 
In the noisy case, this correlation remains, albeit weaker.

\subsection{Likelihood}
We assume a Gaussian likelihood in our analysis,
\begin{align}\label{eq:likelihoods}
\ln \mathcal{L} \propto -\frac{1}{2}\sum_{i,j,X,Y} \left(\hat{S}^{X}_i-\bar{S}^{X}_i \right){\Sigma_{S}^{-1}}^{XY}_{ij} \left(\hat{S}^{Y}_j-\bar{S}^{Y}_j \right),
\end{align}
where $S_i^{X}$ is the $i^\mathrm{th}$ bin of the observable $S$ in redshift bin $X$, $\bar{S}^{X}$ is the mean value at the fiducial cosmology, and $\Sigma_{S}^{XY}$ is the full covariance matrix. 

The Gaussian likelihood is an approximation for the peak and minimum counts as they both follow Poisson distributions. However, since we only consider bins with average counts more than $1.5$ for our $12.25 \, \mathrm{deg}^2$ maps, we expect the number counts for LSST ($\approx 2 \times 10^4 \, \mathrm{deg}^2$) will be sufficiently large to be approximated with a Gaussian distribution.

In order to evaluate the statistics at arbitrary cosmologies, we construct an emulator from the 101 simulated cosmologies using Gaussian Processes~\cite{heitmann2009ii,Kwan2013,Kwan2015,Heitmann2016,Nishimichi2018,McClintock2019}, with the triaxial squared exponential kernel. We can then interpolate the statistics to any cosmology within the sample parameter ranges.   
We verify that the interpolation errors are significantly smaller than the measurement errors. Further details of our emulator are described in detail in \cite{Li2019,Coulton2019}. 
The simulation sample parameter ranges are used as the limits of our flat priors,
\begin{align}
    P\left( \sum m_\nu \right)= &
\begin{cases}
    \text{const},   & \text{if $0\, \mathrm{eV} \leq \sum m_\nu \leq 0.62\,\mathrm{eV}$} \\
    0,              & \text{otherwise}
\end{cases}\\
    P\left( \Omega_m \right) = &
\begin{cases}
    \text{const},   & \text{if $0.18 \leq \Omega_m \leq 0.42$} \\
    0,              & \text{otherwise}
\end{cases}\\
    P\left(A_s\right)=&
\begin{cases}
    \text{const},   & \text{if $1.29 \times 10^{-9} \leq A_s < 2.91 \times 10^{-9}$}\\
    0,              & \text{otherwise}.
\end{cases}
\end{align} 
We then sample the likelihood with a Markov chain Monte Carlo~\citep{foremanmackey2013}. 

\subsection{Parameter constraints}\label{sec:baseConstraints}

\begin{figure*}
  \centering
    \includegraphics[width=.8\textwidth]{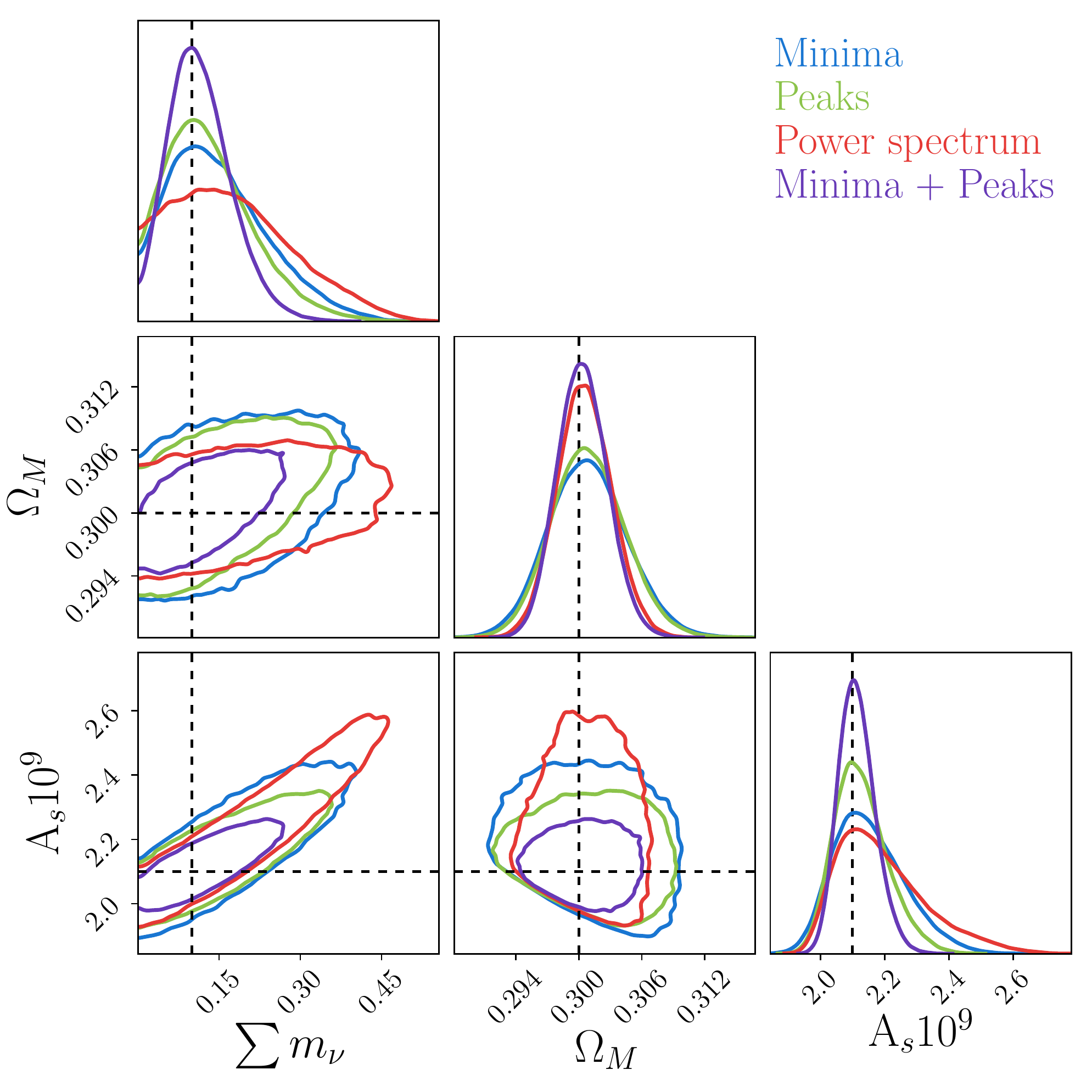}
\caption{$95\%$ credible regions on $\sum m_\nu$, $\Omega_m$, and $A_s$ from the lensing minimum counts, peak counts, and power spectrum, and the combination of minimum and peak counts. Convergence maps are smoothed with a $2\, \mathrm{arcmin}$ Gaussian filter. We use 5 tomographic source redshifts and assume galaxy noises and sky coverage for an LSST-like experiment.}
\label{fig:constraints}
\end{figure*}

In Fig.~\ref{fig:constraints} we show the $95\%$ credible regions on $\sum m_\nu$, $\Omega_m$, and $A_s$ from minimum counts for an LSST-like survey. Convergence maps are smoothed with a $2\, \mathrm{arcmin}$ Gaussian filter. For comparison, we also plot the constraints from the peak counts and power spectrum. For the power spectrum we use $\ell_{\rm max} =3000$. Similar to the peaks, minima are stronger in constraining neutrino mass than the power spectrum. As the power spectrum optimally captures the information of a Gaussian field, the tighter  constraints from minima imply that they are probing non-Gaussian information beyond the second order that is highly sensitive to cosmology. We show results from additional smoothing scales in Appendix~\ref{sec:smoothScales}.


Finally, we explore the joint constraints combining minimum counts with peak counts. We find significant improvement with the joint constraint. 
This shows that minimum counts contains information independent of peak counts. The combined constraints from minimum and peak counts are $44\%$, $11\%$, and $63\%$ tighter than the power spectrum constraints for $\sum m_\nu$, $\Omega_m$, and $A_s$, respectively. 

\section{Impact of baryons}\label{sec:baryon}

\begin{figure*}
  \centering
    \includegraphics[width=.8\textwidth]{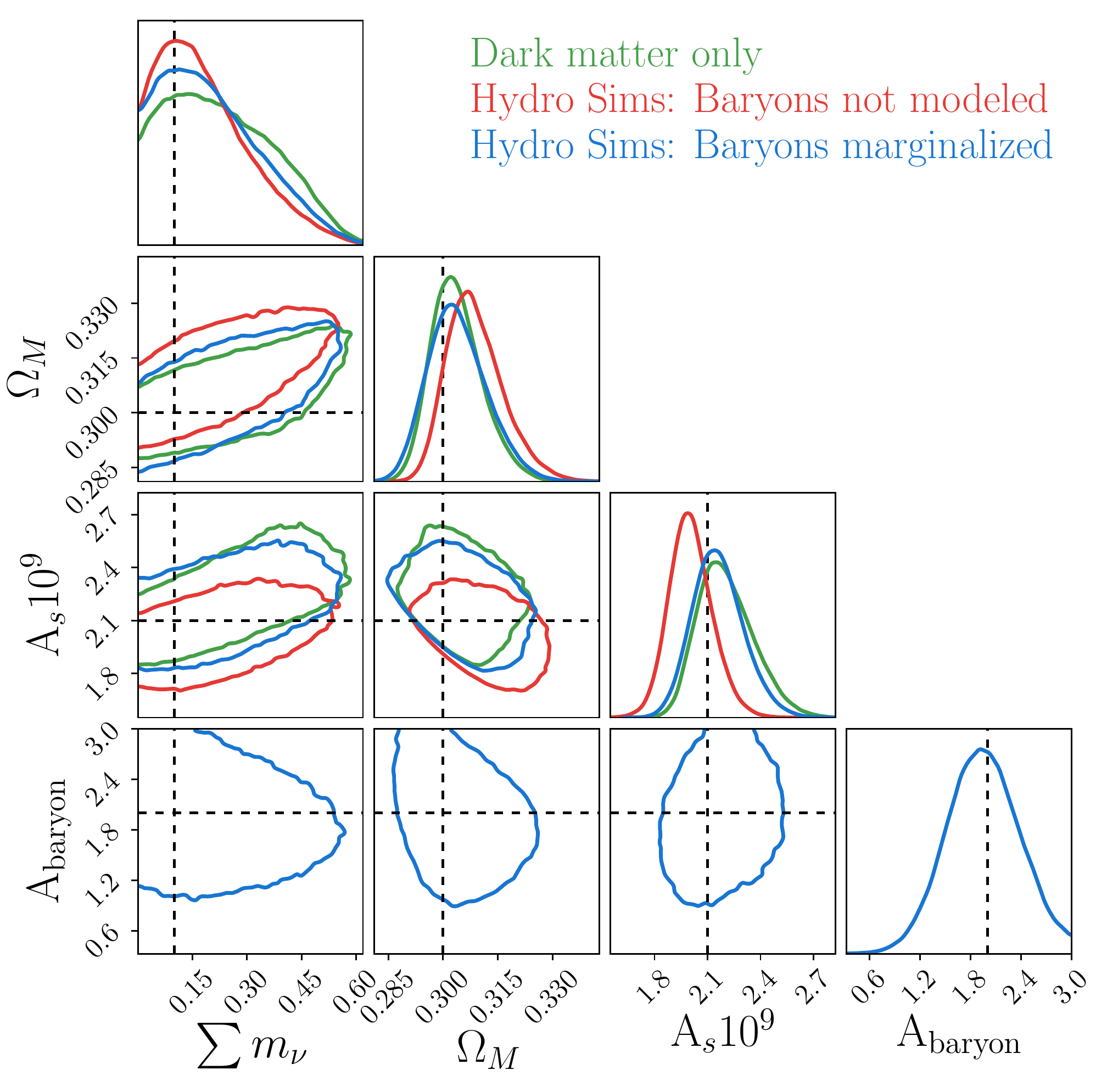}
\caption{$95\%$ credible regions on $\Omega_m$, $A_s$, $\sum m_\nu$ and $A_{\rm baryon}$ for {\bf minimum counts}, for three cases: (1) dark matter-only simulations ({\bf green}), (2) hydrodynamical simulations with dark matter-only models, which would interpret baryonic effects as biases in cosmological parameters (``hydro sims: baryons not modeled'', {\bf red}), and (3) hydrodynamical simulations with models that include baryonic effects (``hydro sims: baryons marginalized'', {\bf blue}), parameterized as $A_{\rm baryon}$. }
\label{fig:constraintsMinimaWithBaryons}
\end{figure*}

We use the \texttt{BAHAMAS} and \texttt{Osato15} simulation suites to explore the impact of baryonic effects on minimum counts and their parameter constraints, using one source redshift\footnote{The effect of using tomography for minimum counts is similar to the effect on peaks \citep[see e.g.][]{Li2019}, which can be seen by comparing the constraints in Figs.~\ref{fig:constraints} and \ref{fig:constraintsMinimaWithBaryons}.} $z_s = 1.0$. While our paper focuses on minimum counts, we also show results on peak counts. All previous works studying baryon effects on peaks either remained at the observable level, or propagated to cosmological constraints but used only a simple Fisher formalism. We present the first study using a large set of cosmologies and full likelihood analysis. 

\begin{figure*}
  \centering
    \includegraphics[width=.95\textwidth]{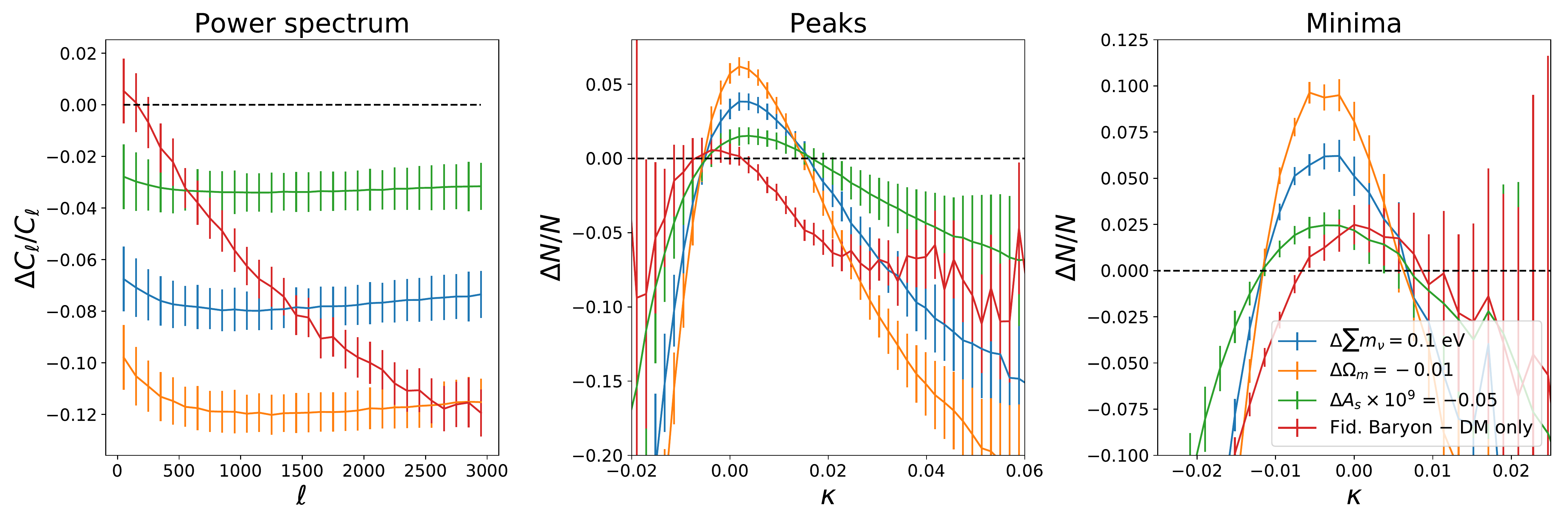}
\caption{The fractional difference in the power spectrum (left), peak counts (center), and minimum counts (right) due to variation of three cosmological parameters ($\sum m_\nu$, $\Omega_m$, $A_s$) and baryons \textbf{without galaxy shape noise}, for a single source redshift $z_s = 1$.  For the baryonic effects, we use the fiducial AGN feedback model from the \texttt{BAHAMAS} simulations.}
\label{fig:effectOfNeutrinoMass}
\end{figure*}

In Fig.~\ref{fig:effectOfNeutrinoMass} we show the fractional difference between the \texttt{BAHAMAS} fiducial AGN model and the corresponding dark matter-only simulations. To see the physics more transparently, we show the effects without the galaxy shape noise. The main effect of baryons on the power spectrum is to suppress the power from $\ell\approx$ a few $\times 100$, consistent with results from previous work \citep[e.g.][]{Gouin2019}. Baryon effects suppress positive and deep negative minima, while boosting the $\kappa\approx$0 minima.  
~\cite{Paillas2017} found evidence in the EAGLE hydrodynamical simulations that baryonic processes can reduce the number of voids and produce less empty voids, both of which could be relevant to our findings. When noise is added, we see similar effects (Fig.~\ref{fig:effectOfNeutrinoMassWithNoise}).  

\begin{figure*}
  \centering
    \includegraphics[width=.8\textwidth]{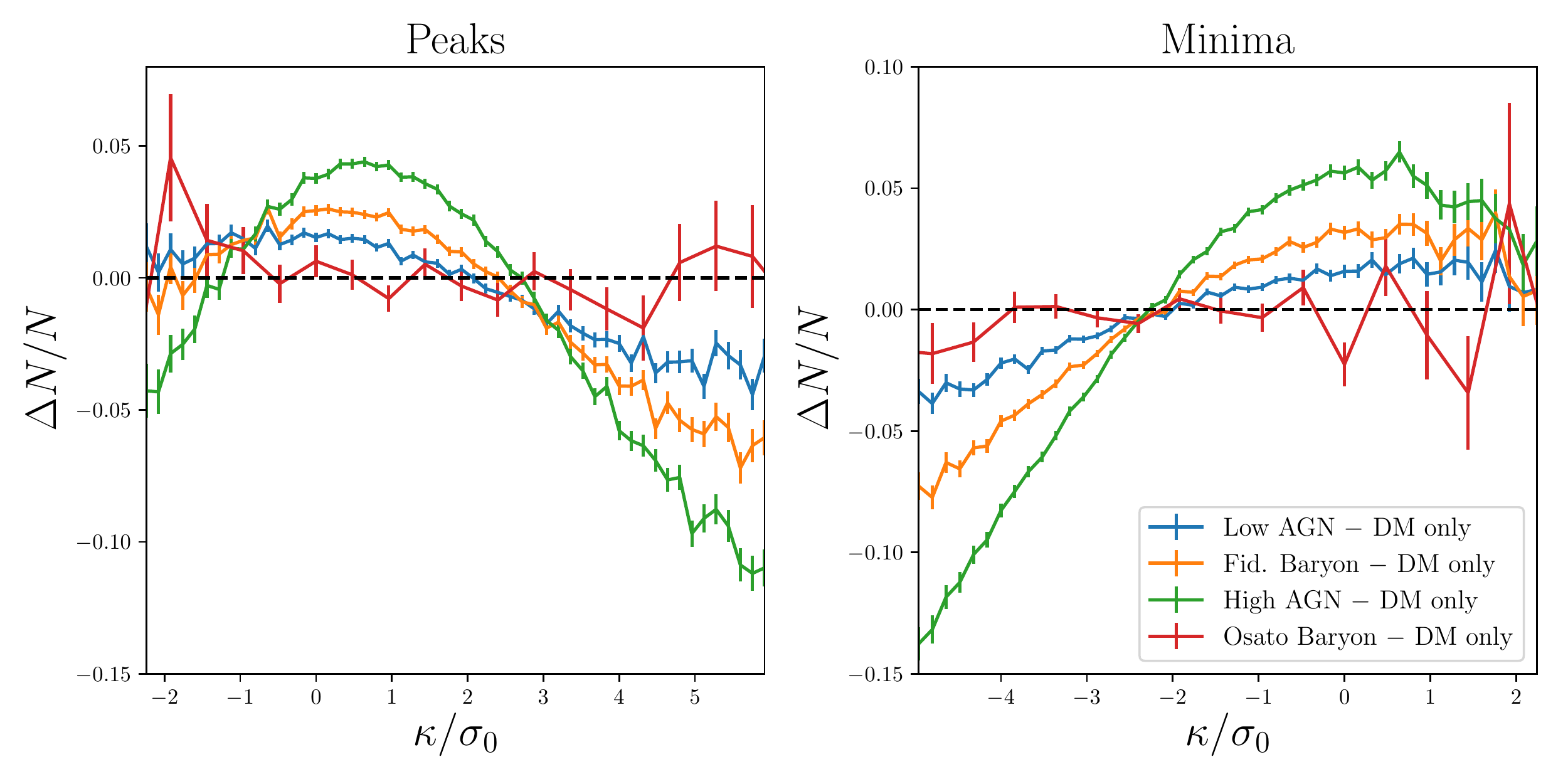}
\caption{The fractional difference in minimum counts and peak counts due to different feedback implementations and strengths. The error bars represent the uncertainties due to the finite number of simulations; thus are given by the measured variance in the simulations $\sigma^\mathrm{measured}_\mathrm{counts}$ divided by the square root of the number of simulations $\sqrt{N_{\rm sim}}$. }
\label{fig:feedbackImps}
\end{figure*}

We also compare the effects of baryons to those from changing cosmological parameters in Fig.~\ref{fig:effectOfNeutrinoMass}. One promising feature is that the fractional effects on all three observables are different for baryons than for cosmological parameters investigated here. However, effects of baryons rely heavily on subgrid models in hydrodynamical simulations, which remain somewhat uncertain~\cite{Springel2018}. 
This is demonstrated in Fig.~\ref{fig:feedbackImps} where we show the fractional effect of baryons for the two other AGN feedback models in the \texttt{BAHAMAS} simulations and the ``FE'' model from \texttt{Osato15}.  However, we note that \texttt{BAHAMAS} simulations are calibrated to match some key observables, such as the present-day baryon content of massive systems, the hot gas mass fraction--halo mass relation of galaxy groups and clusters, as well as the amplitude of the black hole mass--stellar mass relation. Hence we expect them to be more realistic and use them to study the impact of baryons on cosmological constraints. Further, the \texttt{BAHAMAS} low and high AGN  models were constructed to capture the upper and lower bounds of the observed group and cluster gas fractions, and hence we expect them to represent the theoretical uncertainties in the effects of baryons, making them ideal to study the impact of baryons on cosmological constraints.

\begin{figure*}
  \centering
    \includegraphics[width=.8\textwidth]{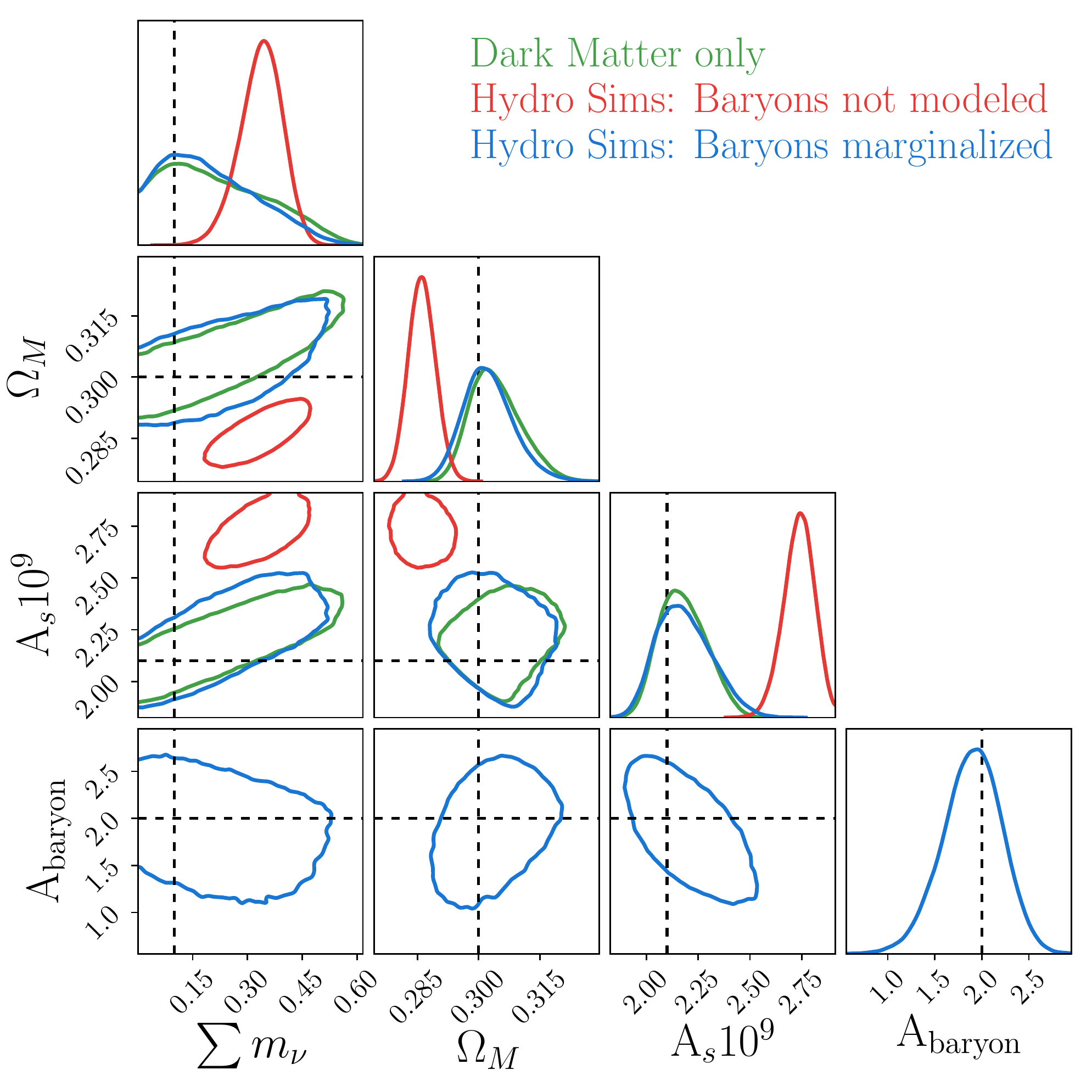}
\caption{$95\%$ credible regions on $\Omega_m$, $A_s$, $\sum m_\nu$ and $A_\mathrm{baryon}$ for \textbf{peak counts}, for three cases: (1) dark matter-only simulations (\textbf{green}), (2) hydrodynamical simulations with dark matter-only models, which would interpret baryonic effects as biases in cosmological parameters (``hydro sims: baryons not modeled'', \textbf{red}), and (3) hydrodynamical simulations with models that include baryonic effects (``hydro sims: baryons marginalized'', \textbf{blue}), parameterized as $A_\mathrm{baryon}$. }
\label{fig:constraintsPeaksWithBaryons}
\end{figure*}

Next, we use the \texttt{BAHAMAS} simulations to model the impact of baryons on cosmological constraints. 
We make the assumption that the fractional effect of baryonic processes is independent of cosmology, which has been shown to be true for the power spectrum within a few percent~\cite{Mead2015,Mead2016,Mummery2017,vanDaalen2019,Stafford2019}. 
%
%
To include baryonic effects, we introduce into our emulator a new parameter $A_\mathrm{baryon}$, which linearly interpolates the fractional effect of baryons between no baryonic effects ($A_\mathrm{baryon} = 0$) to the \texttt{BAHAMAS} high-AGN model $A_\mathrm{baryon} = 3$, with $A_\mathrm{baryon} = 2$ for the fiducial model and $A_\mathrm{baryon} = 1$ for the low-AGN model. We then jointly fit $A_\mathrm{baryon}$ together with the cosmological parameters. 

We show our results with baryons for lensing minima in Fig.~\ref{fig:constraintsMinimaWithBaryons} 
and lensing peaks in Fig.~\ref{fig:constraintsPeaksWithBaryons}.
We compare three cases: (1) ``dark matter only'' simulations, (2) ``hydro sims: baryons not modeled'': hydrodynamical simulations as the observable, but fitted to dark matter-only models, i.e.\,using an emulator with only three cosmological parameters ($\sum m_\nu$, $\Omega_m$, $A_s$). This method would interpret baryonic effects as biases in cosmological parameters, and (3) ``hydro sims: baryons marginalized'': hydrodynamical simulations with models that include baryonic effects with four parameters ($\sum m_\nu$, $\Omega_m$, $A_s$, $A_\mathrm{baryon}$).

Without modeling baryons, cosmological constraints from lensing minima have mild biases $\approx 0.5 \sigma$ from the true values. In contrast, we see significant biases using peak counts, at more than $4 \sigma$ from the true values. This implies that minimum counts are a more robust statistic against baryonic effects than the peak counts. However, when we marginalize the baryonic effects (case 3), the biases vanish for both minimum counts and peak counts, implying that the effects of baryons are sufficiently distinct from the cosmological parameters and we can mitigate them by including baryons in our model.

\section{Conclusions}\label{sec:conclusions}
We study the cosmological constraints from weak lensing minima, a simple statistic complementary to the lensing power spectrum and peak counts. Our analysis used realistic galaxy redshift distribution and shape noise as expected from LSST. We find that they contain non-Gaussian information, and provide tighter constraints than the power spectrum. Lensing minima alone are slightly less constraining than the peaks. However, when the two are combined, they produce significantly tighter constraints than the power spectrum, by $44\%$, $11\%$, and $63\%$ on $\sum m_\nu$, $\Omega_m$, and $A_s$, respectively. Our results show that lensing minima are a promising probe for upcoming cosmological experiments.

We use hydrodynamical simulations to study the effects of baryons on lensing minima. We find that baryonic processes result in reduced number of deep (very negative $\kappa$) and high, positive $\kappa$ minima, while enhancing the number of shallower ($\kappa \approx 0$) minima. We find that the baryonic effects, as modeled in the hydrodynamical simulations \texttt{BAHAMAS}, have little impact on minimum counts ($\approx 0.5 \sigma$ biases), but can induce large ($\ge 4 \sigma$) biases in peak counts analysis. By extending our emulator to include baryonic processes and marginalize them, we recover the correct cosmology without losing much constraining power for both the minimum counts and peak counts. Our results emphasize the importance of modeling baryonic effects for future lensing surveys, and suggest that lensing minima can be a useful tool to mitigate the biases induced by baryons.

Our work is the first step to investigate the cosmological constraints from lensing minima and the effects of baryons on them. Future work should study other systematics, such as the intrinsic alignments of galaxies, photometric redshifts, multiplicative biases in shape measurements. As we used a simple linear model to capture the baryonic effects, more accurate modeling of baryons will be beneficial, such as a more general parameterization~\citep{HarnoisDeraps2015,Schneider2019} or via a Principal Component Analysis \citep{Eifler2015,Huang2019}. 
\section{Acknowledgements}

We thank Cora Uhlemann, Oliver Friedrich, Daan Meerburg, and Patrick de Perio for helpful discussions.
IGM thanks Joop Schaye for his contributions to the \texttt{BAHAMAS} project.
This work is supported by an NSF Astronomy and Astrophysics Postdoctoral Fellowship (to JL) under award AST-1602663. We acknowledge support from the WFIRST project. WRC acknowledges support from the UK Science and Technology Facilities Council (grant number ST/N000927/1).
This project has received funding from the European Research Council (ERC) under the European Union's Horizon 2020 research and innovation programme (grant agreement No 769130).
KO is supported by Japan Society for the Promotion of Science (JSPS) Overseas Research Fellowships.
We thank New Mexico State University (USA) and Instituto de Astrofisica de Andalucia CSIC (Spain) for hosting the Skies \& Universes site for cosmological simulation products. This work used resources of the National Energy Research Scientific Computing Center (NERSC), a U.S. Department of Energy Office of Science User Facility operated under Contract No.\,DE-AC02-05CH11231. 
This work used the Extreme Science and Engineering Discovery Environment (XSEDE), which is supported by NSF grant ACI-1053575. The analysis is in part performed at the TIGRESS high performance computer center at Princeton University.
This work used the DiRAC@Durham facility managed by the Institute for Computational Cosmology on behalf of the STFC DiRAC HPC Facility (\url{https://www.dirac.ac.uk}). The equipment was funded by BEIS capital funding
via STFC capital grants ST/K00042X/1, ST/P002293/1, ST/R002371/1 and
ST/S002502/1, Durham University and STFC operations grant
ST/R000832/1. DiRAC is part of the National e-Infrastructure.
Numerical simulations were in part carried out on Cray XC30 at the Center for Computational Astrophysics,
National Astronomical Observatory of Japan.

\appendix
\section{The effect of smoothing scales}\label{sec:smoothScales}

\begin{figure*}
  \centering
    \includegraphics[width=.8\textwidth]{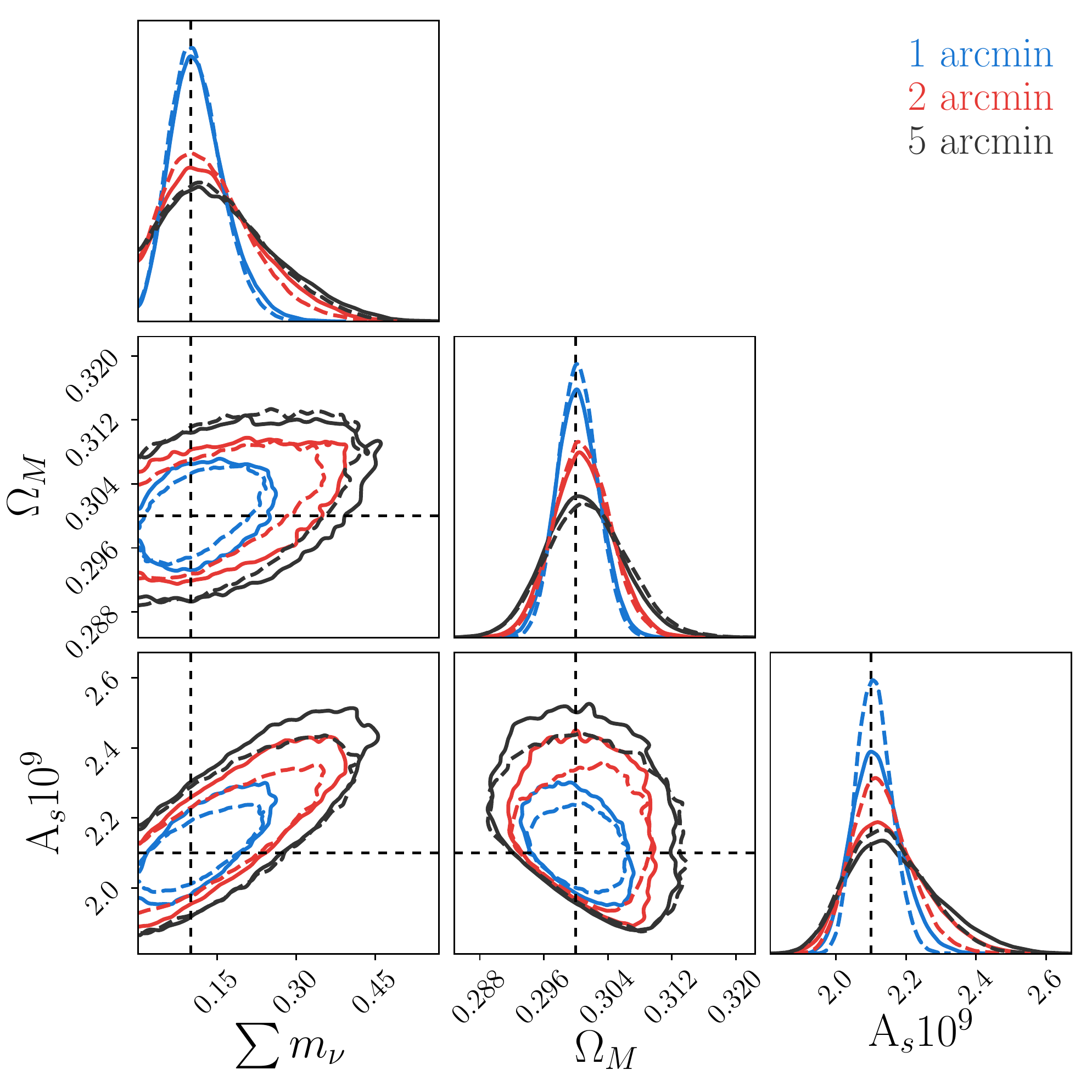}
\caption{$95\%$ credible regions from minimum counts ({\bf solid lines}) and peak counts ({\bf dashed lines}) for three different smoothing scales, using 5 tomographic redshifts and LSST-like galaxy noise.}
\label{fig:constraintsVsSmoothing}
\end{figure*}

In the main analysis, we used a $2 \, \mathrm{arcmin}$ smoothing scale. Here we explore how the $95\%$ credible regions from lensing minima and peaks change as we vary the smoothing scales ($1$, $2$, and $5 \, \mathrm{arcmin}$). In Fig.~\ref{fig:constraintsVsSmoothing}, we find that the constraining power degrades quickly with increasing smoothing scales for both statistics, as we start to lose small scales where non-Gaussian information is the richest. 
Most past works on cosmological constraints from weak lensing data used large smoothing scales ($\ge10 \, \mathrm{arcmin}$) to avoid baryonic effects. From our main results, we find that lensing peaks can indeed be highly biased by baryonic effects and hence requires careful modeling. However, we find lensing minima somewhat insensitive to baryonic effects and hence may be a more robust tool to calibrate cosmology on small scales.  

\section{Comparison to genus}\label{sec:minkFuncs}

\begin{figure*}
  \centering
    \includegraphics[width=.8\textwidth]{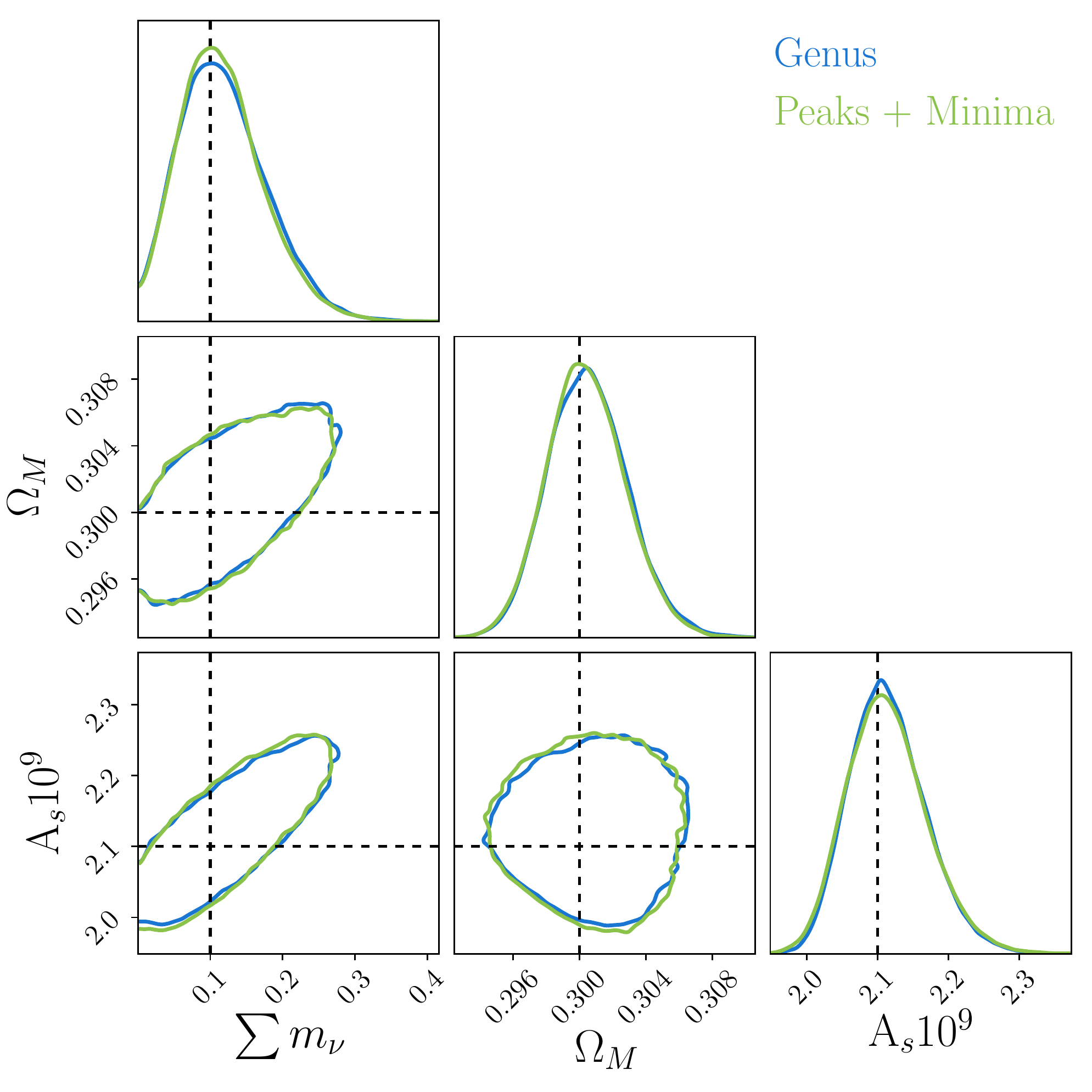}
\caption{$95\%$ credible regions from the combination of minimum counts and peak counts (green) and that from the third Minkowski functional the genus (blue), using 5 tomographic redshifts and LSST-like galaxy noise.}
\label{fig:constraintsVsMinkFuncs}
\end{figure*}

Minkowski functionals characterize the morphology of a field and are sensitive to non-Gaussian distributions. For a two dimensional field, there are three Minkowski functionals: the area $V_0$,   perimeter $V_1$, and genus $V_2$. Previous work by \cite{Petri2013,Petri2015,Marques2019} found that Minkowski functionals can offer strong cosmological constraints. In this Appendix, we are particularly interested in the genus $V_2$ --- the difference between the number of ``holes'' and the number of ``islands'', which we think can be closely related to the minimum counts and peaks counts in the field, respectively. 

Genus is expressed as a function of the threshold $\kappa_0$,
\begin{align}
V_2(\kappa_0) = \frac{1}{2\pi A}\int\limits_{\partial\Sigma(\kappa_0)}\mathrm{d}l \mathcal{K}, 
\end{align}
where 
$A$ is the total area of the field, 
$\mathcal{K}$ is the curvature, $\Sigma(\kappa_0)$ is the excursion set of all pixels with $\kappa\ge \kappa_0$, and $\partial\Sigma(\kappa_0)$ is the boundary of the excursion set. 
Their connection can be explicitly verified by examining their cosmological information. In Fig.~\ref{fig:constraintsVsMinkFuncs} we show the parameter constraints from the genus of the lensing field and the combination of lensing peaks and minima. We find that their constraints are almost identical.

\bibliographystyle{mnras}
\interlinepenalty=10000
\bibliography{project}
\bsp	
\label{lastpage}
\end{document}